\documentclass[reprint, superscriptaddress, amsmath, field, amssymb, aps]{revtex4-2}
\usepackage{xcolor}
\usepackage[version=4]{mhchem}
\usepackage{lipsum} 
\usepackage[version=4]{mhchem}

\usepackage{appendix}
\usepackage{braket}
\usepackage{chemformula}
\usepackage{graphicx}
\usepackage{dcolumn}
\usepackage{bm}
\usepackage{setspace}
\usepackage{physics}
\usepackage{braket}
\usepackage{graphicx}
\usepackage{dcolumn}
\usepackage{bm}
\usepackage[colorlinks=true, citecolor=blue, linkcolor=blue, urlcolor=blue]{hyperref}
\usepackage{indentfirst}
\usepackage{bbm}
\usepackage[english]{babel}

\usepackage[thinc]{esdiff}

\def \ETH{Institute for Theoretical Physics, ETH Z\"urich, CH-8093 Z\"urich, Switzerland}
\def \Harvard{Lyman Laboratory, Department of Physics, Harvard University, Cambridge, MA 02138, USA}
\def \Stanforda{Department of Applied Physics, Stanford University, Stanford, California 94305, USA}
\def \Stanfordb{E. L. Ginzton Laboratory, Stanford University, Stanford, California 94305, USA}
\def \Stanfordc{Department of Physics, Stanford University, Stanford, California 94305, USA}
\def \Standrews{SUPA, School of Physics and Astronomy, University of St. Andrews, St. Andrews KY16 9SS, United Kingdom}

\begin{document}

\title{Superradiant Charge Density Waves \\in a Driven Cavity-Matter Hybrid}
\author{Luka Skolc}
\thanks{These authors contributed equally to this work}
\affiliation{\ETH}

\author{Sambuddha Chattopadhyay}
\thanks{These authors contributed equally to this work}
\affiliation{\ETH}
\affiliation{\Harvard}

\author{Filip Marijanović}
\affiliation{\ETH}

\author{Qitong Li}
\affiliation{\Stanforda}
\affiliation{\Stanfordb}

\author{Jonathan Keeling}
\affiliation{\Standrews}

\author{Benjamin L. Lev}
\affiliation{\Stanforda}
\affiliation{\Stanfordb}
\affiliation{\Stanfordc}

\author{Eugene Demler}
\affiliation{\ETH}

\date{\today}
\begin{abstract}
Optical cavities enable strong, long-range, light–matter interactions that can drive collective ordering phenomena, such as superradiant self-organization in ultracold atomic gases. Extending these ideas to solid-state electron systems could enable continuous-wave optical control of electronic order, but is impeded by the mismatch between optical wavelengths and electronic length scales. Here, we propose a platform for realizing \textit{superradiant charge density waves} (sCDWs) in doped, driven transition-metal dichalcogenides coupled to an optical cavity. A nanoscale grating generates electric fields at large in-plane optical momenta, allowing cavity photons to couple efficiently to electronic density fluctuations through exciton–polaron processes. Using a linear-stability analysis, we determine the threshold for superradiant ordering and map out the driven phase diagram. We show that tuning the grating periodicity to match the enhanced electronic density fluctuations---such as those near Wigner crystallization---substantially lowers the required pump intensity. Our results establish a novel route toward cavity-controlled electronic order in quantum materials.
\end{abstract}

\maketitle
The optical control of quantum materials is a focal project of contemporary solid-state physics. Promising experiments have demonstrated photocontrol over magnetism, topology, and correlated electronic properties \cite{basov_towards_2017,de_la_torre_colloquium_2021}. Yet the most striking light-induced phenomena---Floquet-Bloch states~\cite{Flouqet1, Flouqet2, Flouqet3, Flouqet4, Flouqet5}, light-induced anomalous Hall effect~\cite{McIver}, photoinduced superconductivity~\cite{sc1, sc2, sc3, sc4, sc5, sc6, sc7, sc8, sc9, sc10}---often require ultrashort pulses with giant fields and vanish once heating sets in. Achieving continuous-wave optical control is therefore crucial, both for enabling applications and for characterizing light-induced electronic states with conventional probes such as transport or STM. 
 
A promising route to continuous-wave optical control is to combine low-power laser driving with the strong light–matter interaction enhancement provided by optical cavities~\cite{Bourzutschky2024rci}. Pioneering ultracold atom experiments have demonstrated this paradigm: Under strong driving, a homogeneous quantum gas coupled to a cavity self-organizes into a density wave that acts as a Bragg grating for the incident laser light, leading to a superradiant buildup of cavity photons~\cite{coldAtoms1, coldAtoms2, coldAtoms3, coldAtoms4, coldAtoms5, coldAtoms6}. However, translating this mechanism to solid-state platforms faces two key challenges. First, electronic length scales---unlike the interatomic spacings in ultracold gases---are orders of magnitude smaller than optical wavelengths. Consequently, cavity photons impart negligible momentum to electrons, which strongly suppresses density-wave formation. Second, the microscopic mechanism that couples atomic density waves to the optical cavity in ultracold atom experiments involves a Raman scheme between hyperfine atomic levels where one optical transition is mediated by a pump photon and the other by a cavity photon. The quantum-optical interference between these two pathways induces the atomic density-wave order observed at sufficiently strong transverse driving. However, unlike atoms, electrons lack optically addressable internal structure, complicating further the transcription from atoms to solids. 

In this work, we discuss an experimentally realizable solution to both of these problems. By synthesizing concepts from ultracold atoms, nanophotonics, and solid-state physics, we propose a presently feasible route towards the continuous-wave optical control via superradiant self-organization in a driven cavity.  
We use the term \textit{superradiant charge density waves} (sCDWs) to denote the charge-density-wave order that we predict may be photo-induced in the solid state. We propose that sCDWs can be realized in lightly doped transition-metal dichalcogenides (TMDs) placed in an optical cavity~\cite{confocal1, confocal2} and transversely pumped with laser light. 
\begin{figure}
        \centering
        \includegraphics[width = 0.999 \linewidth]{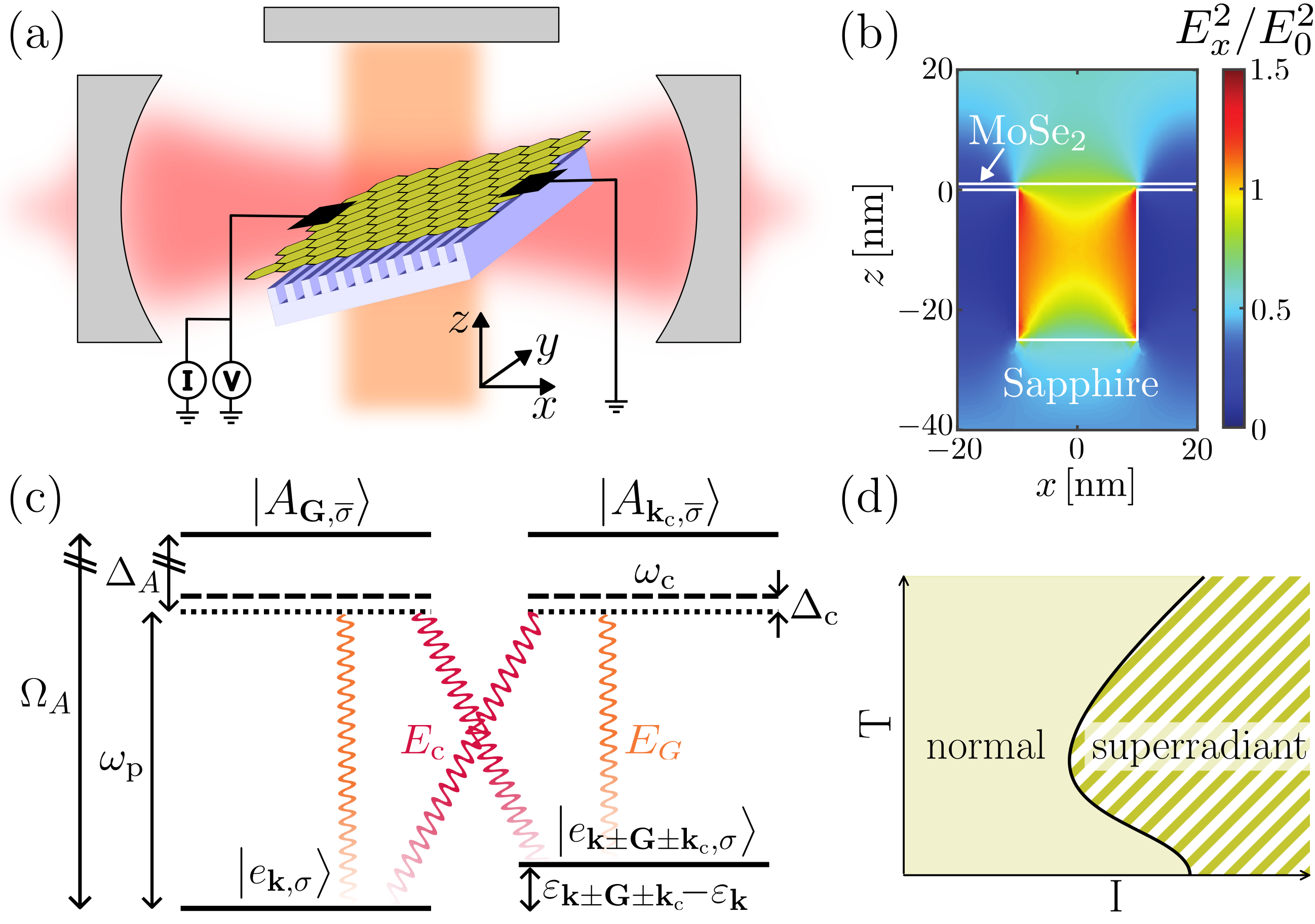}
        \caption{\textbf{Proposed experimental setup.} (a) Monolayer TMD on a grated substrate illuminated by a pump laser (orange) and interacting with an optical cavity mode (red). The sample is tilted for optimal coupling to the cavity and leads can be unobstructively attached to perform transport measurements. (b) Simulated electric field close to one repeat unit of the grating. $E_x$ is the electric field along $x$ and $E_0$ is the pump field far away from the grating. The thin horizontal line above the grating boundary (staggered line) shows the position of the encapsulated TMD sample. (c) Level diagram for the proposed light-matter coupling scheme. Pump and cavity fields can excite or annihilate excitons that combine with electrons $\ket{e}$ from opposite valleys to form attractive polarons $\ket{A}$. The pump frequency is closely detuned to the cavity mode and the attractive polaron branch. (d) Schematic phase diagram as a function of pump laser intensity and temperature: in the normal phase, the density of mobile electrons is uniform. Above a critical pump intensity, the cavity superradiates (the cavity mode acquires a coherent displacement) while the electrons order into stripes.}
        \label{fig:setup}
\end{figure}

Our proposed solution to the significant mismatch between electronic (${\sim} 10{\rm s} ~{\rm nm}$) and optical (${\sim} 1~\mu{\rm m}$) length scales is to place the doped TMDs on a substrate that is etched with a period corresponding to the ${\sim} 10{\rm s}~{\rm nm}$ electronic length scale; see Fig.~\ref{fig:setup}(a). This etching serves as a grating, converting a large fraction of the incident pump light into a subwavelength lattice at \textit{optical frequencies} with \textit{electronic wave-numbers}, allowing for efficient coupling of photons to electrons. Regarding the second problem, we engineer a Raman scheme with electrons by tuning our pump and cavity to be near-resonant with sharp exciton-polarons in the 2D TMDs; see Fig.~\ref{fig:setup}(b) and also~\cite{Bourzutschky2024rci}. By combining these two approaches, we can extend the concept of superradiant density order from ultracold atoms to the solid state. Leveraging a linear-stability analysis, we map out the driven phase diagram and show that sCDWs, arising from the all-to-all coupling introduced by coupling electronic density to an optically driven cavity, can be realized with reasonable continuous-wave laser intensities in the driven-cavity setting.

\section{Results}
\subsection{Proposed Experimental Setup}
A two-dimensional monolayer TMD material encapsulated in hBN is placed in a transversely pumped optical cavity. The TMD sits on top of a sapphire substrate coated with an anti-reflection (AR) coating; Fig.~\ref{fig:setup}(a). This top coating is etched to form a grating with a period $\lambda\approx 10-40~\rm nm$, which is comparable to electronic length scales in dilutely doped 2D semiconductors ($\nu_{\rm e}\lesssim 10^{12}~\mathrm{cm}^{-2}$) \cite{sidler2017fermi}. When illuminated by a near-infrared pump laser, the grating generates high-momentum components in the near-field; Fig.~\ref{fig:setup}(b). In essence, the grating allows us to  produce optical-frequency light at electronic length scales. With an optimized choice of \ch{Ta2O5} as the top layer of the AR coating, around $12 \%$ of the incident pump light can be converted into a high-momentum component in the near field (see Methods and finite-element simulations in Appendix B of the Supplementary Information). 

\subsection{Microscopic model}
We consider itinerant valley electrons interacting with optically bright excitons in a doped 2D monolayer illuminated by pump and cavity fields. The electron-exciton-photon system is described by
\begin{align}
    H = \,&H_{\rm e} + H_{e-x}  +\sum_\mathbf{k} \Omega_\mathbf{k} X_{\mathbf{k},\sigma}^\dag X_{\mathbf{k},\sigma} +\nonumber\\
    & + d_x \sum_{\mathbf{k},\sigma} E_\mathbf{k} (X_{-\mathbf{k},\sigma} + X_{\mathbf{k},\sigma}^\dag)+ \omega_{\rm c} a^\dag a,\label{eq:H_full}
\end{align}
with $H_e\!=\!\sum\limits_{\mathbf{p}, \sigma} \varepsilon^\sigma_\mathbf{p} c^\dag_{\mathbf{p}, \sigma} c_{\mathbf{p}, \sigma} + \!\!\!\!\sum \limits_{\substack{\mathbf{p}, \mathbf{p}', \mathbf{q},\sigma,\sigma'}} \!\!\!\!\!\!\! V_{\mathbf{q}} c^\dag_{\mathbf{p} + \mathbf{q},\sigma} c^\dag_{\mathbf{p}'- \mathbf{q}, \sigma'} c_{\mathbf{p}', \sigma'} c_{\mathbf{p},\sigma}$ describing interacting valley electrons and $H_{e-x}\! =\! \sum \limits_{\substack{\mathbf{p}, \mathbf{p}', \mathbf{q}, \sigma, \sigma'}} \!   U^{\sigma,\sigma'}_{\mathbf{q}} X^\dag_{\mathbf{p} + \mathbf{q},\sigma} c^\dag_{\mathbf{p}' -\mathbf{q}, \sigma'} c_{\mathbf{p}', \sigma'} X_{\mathbf{p}, \sigma}$ capturing the electron-exciton interactions \cite{sidler2017fermi,wang2025spectroscopywignercrystalpolarons}. The operator $c^\dag_{\mathbf{k},\sigma}$ creates an electron in the valley $\sigma$ with momentum $\mathbf{k}$, $X^\dag_{\mathbf{k},\sigma}$ creates an intravalley exciton with total momentum $\mathbf{k}$, and $\Omega_{\mathbf{k}}$ is the exciton dispersion. The first three terms in Eq.~\eqref{eq:H_full} give rise to three-body (bound) states of excitons and electron-hole pairs known as exciton polarons. We focus specifically on attractive exciton polarons, treating them within the commonly used Chevy ansatz (see Methods) ~\cite{sidler2017fermi}.

The light-matter coupling term involves the dipole operator $d_{x}$ and the total electric field incident on the TMD monolayer $E_\mathbf{k} = E_{\rm c} (a+ a^\dag) \left( \delta_{\mathbf{k}, \mathbf{k}_{\rm c}} + \delta_{\mathbf{k}, -\mathbf{k}_{\rm c}} \right) + E_G \left(e^{-i\omega_{\rm p} t} + e^{i\omega_{\rm p} t}\right) \left( \delta_{\mathbf{k}, \mathbf{G}} + \delta_{\mathbf{k},-\mathbf{G}} \right) $, which is a combination of the cavity field and the strong, pumped, grating-modified evanescent near field. The total electric field is characterized by the pump frequency $\omega_{\rm p}$, the grating wavevector $\mathbf{G} = \frac{2\pi}{\lambda}\hat{e}_x$, the momentum-space component of the pump electric field strength at the wavevector $E_G$ of the first-order spatial mode of the modulated evanescent field, the strength of cavity field fluctuations $E_{\rm c}=\sqrt{\frac{\hbar\omega_{\rm c}}{2\epsilon_0 \epsilon V_{\rm c}}}$, the cavity mode volume $V_{\rm c}$, the dielectric permittivity of the encapsulating hBN $\epsilon$, the cavity wavevector $k_{\rm c}\ll G$, the frequency of the cavity mode $\omega_{\rm c}$ and the operator $a^{\dag}$ that creates a cavity photon.

\subsection{Effective model for light-matter coupling}
Starting from Eq.~\eqref{eq:H_full}, we derive an effective Hamiltonian for electron-photon interactions mediated by virtual exciton-polaron excitations (see Methods). We irradiate the system with light that is near-resonant with the attractive polaron albeit sufficiently detuned to be in the dispersive regime. By doing so, we achieve strong coupling between electronic density modulations and the cavity photon displacement
\begin{align}
    H_{\rm eff}=&\, H_{\rm e} + \frac{\Delta_{\rm c}}{2} \left(X_{\rm c}^2 + P_{\rm c}^2\right) + \Lambda X_{\rm c} \sum\limits_{\mathbf{Q} = \mathbf{Q}_\pm} \left( \rho_{\mathbf{Q}} + \rho_{-\mathbf{Q}}\right), \label{eq:H_eff}
\end{align}
where $\Delta_c= \omega_c - \omega_p$, $X_{\rm c}=\frac{1}{\sqrt{2}}(a+a^\dag)$ and $P_{\rm c}=\frac{i}{\sqrt{2}}(a^\dag- a)$ are the cavity quadratures, $\mathbf{Q}_\pm = \mathbf{G}\pm \mathbf{k}_{\rm c}$ are the ordering wavevectors and $\rho_\mathbf{Q} = \sum\limits_{\mathbf{p}, \sigma} c^\dag_{\mathbf{p}+\mathbf{Q}, \sigma}c_{\mathbf{p}, \sigma}$ is the electronic density modulation at wavevector $\mathbf{Q}$. The effective coupling strength
\begin{equation}
    \Lambda = d_x^2 E_G E_{\rm c} \frac{8\sqrt{2}\pi a_T^2}{A\Delta_{\rm A}}, \label{eq:lambda}
\end{equation}
where $A$ is the sample area and $a_T$ is the trion binding radius, stems from virtual processes shown in Fig.~\ref{fig:setup}(c)
that effectively introduce a double Lambda scheme for the electrons, similar to that utilized in realizing self-organization of ultracold atoms in optical cavities~\cite{bosonTheo1, coldAtoms2}. $\Delta_A = E_A-\omega_p$ is the detuning of the attractive polarons to the pump and $E_A$ is the energy of the attractive polaron. 

We elaborate here on the microscopic processes underlying the light-matter coupling term in Eq.~\eqref{eq:H_eff}. The pump field scattered by the grating excites an exciton with finite momentum $\mathbf{G}$, which then combines with a hole and an electron from the opposite valley to form an attractive intervalley polaron with total momentum $\mathbf{G}$. The polaron dissociates, emitting a photon into the cavity field and simultaneously annihilating an exciton with momentum $\mathbf{k_{\rm c}}$. In addition to the described process, it is also possible for a polaron with total momentum $\mathbf{k}_{\rm c}$ to be created by absorbing a cavity photon and dissociated by emitting into the pump field. In both cases, the difference of the pump and cavity photon momenta $\pm\mathbf{G}\pm\mathbf{k}_{\rm c}$ is imparted on the Fermi sea, generating the $\rho_{\pm\mathbf{Q}_\pm}$ terms in Eq.~\eqref{eq:H_eff}. The trionic length scale $a_{\rm T} \approx 2 ~{\rm nm}$ enters Eq.~\eqref{eq:lambda} via the attractive polaron wavefunction, since in the dilute limit, the attractive polaron possesses a strong trionic character (see Methods).

In passing, we note that a substantially different approach for exploiting driven cavity superradiance to induce phonon-polariton condensates has been discussed in Ref.~\cite{Bourzutschky2024rci}. Beyond the commonality that both proposals involve driven cavities, they are rather separate: We propose inducing electronic sCDWs, whereas the other experiment proposed inducing a phononic instability. We find, however, that the laser intensities required in our scheme are likely lower, leading to reduced sample heating.

\subsection{Superradiant threshold}
To locate the superradiant phase transition, we perform a linear stability analysis on Eq.~\eqref{eq:H_eff} by linearizing Heisenberg equations of motion around the (possibly unstable) electronic ground state for $\Lambda=0$~\cite{marijanović2024dynamicalinstabilitiesstronglyinteracting}. We determined the eigenmodes of the linearized system by simultaneously solving: $\omega X_{\rm c} =-i\Delta_{\rm c} P_{\rm c}$, $\omega P_{\rm c}=i\Delta_{\rm c} X_{\rm c} + i\Lambda \sum\limits_{\mathbf{Q} = \mathbf{Q}_\pm} (\rho^{\rm e}_\mathbf{Q} + \rho^{\rm e}_{-\mathbf{Q}})$, and $\rho^{\rm e}_\mathbf{Q}+ \rho^{\rm e}_{-\mathbf{Q}}= 2 N_{\rm e} X_{\rm c}\Lambda \chi(\mathbf{Q},\omega)$, where $\chi(\mathbf{Q},\omega)$ is the dynamical electronic density susceptibility and $N_{\rm e}$ is the number of mobile electrons. We derive the self-consistency condition 
\begin{equation}
    \Delta_{\rm c}^2=\omega^2+ 2\Delta_{\rm c} N_{\rm e} \Lambda^2 \sum\limits_{\mathbf{Q} = \mathbf{Q}_\pm} \chi(\mathbf{Q}, \omega),\label{eq:self_consistency}
\end{equation} 
which captures unstable polaritonic modes when a solution for $\omega$ has a positive imaginary part. The onset of superradiance occurs when $\Im(\omega)=0$. If we further assume the instability is not oscillatory (e.g., is not a limit cycle) then the critical condition occurs at $\omega = 0$, such that 
\begin{equation}
    1= 2 \frac{N_{\rm e} \Lambda_c^2}{\Delta_{\rm c}} \sum\limits_{\mathbf{Q} = \mathbf{Q}_\pm} \chi(\mathbf{Q})\label{eq:threshold}
\end{equation}
determines the critical coupling strength $\Lambda_c$, above which electrons develop density order that scatters light between the cavity and pump fields, and the cavity field acquires a finite coherent displacement, thereby superradiating. For $\Lambda<\Lambda_c$, the electron density is uniform and the cavity is in a vacuum state; Fig.~\ref{fig:setup}(d). 

The crucial role of the nanoscale etching becomes manifest here. In the absence of the grating, the relevant density fluctuations occur at optical wavelengths. As such, the vanishingly small electrostatically screened $\chi(\mathbf{Q}\rightarrow 0)$ would increase the threshold laser power to intensities far above the damage threshold. 
Instead, by tuning the grating periodicity $\lambda$ and with it $\mathbf{Q_\pm}$, we can exploit different peaks in $\chi(\mathbf{Q})$ to lower the critical pump intensity in Eq.~\eqref{eq:threshold}. Peaks in $\chi(\mathbf{Q})$ may arise either from proximity to density-ordering phase transitions (e.g., Wigner Crystals, conventional charge density waves) or from Kohn-Lindhard features due to nesting wave-vectors. In the following sections, we explore design principles for efficient sCDW production that exploit enhanced density fluctuations to lower the required laser intensities. 

\subsection{Superradiance close to Wigner crystallization}\label{sec:wigner}
Close to a density-ordering phase transition, density fluctuations in the disordered phase are enhanced at the wavevector of the incipient, unformed crystal. We explore how we can exploit enhanced density fluctuations in the vicinity of the Wigner crystal phase transition, a well-known density-ordering phenomenon in dilute electronic fluids where Coulomb interactions dominate kinetics and a crystalline state spontaneously forms once the electronic density decreases below a critical value. 

\begin{figure}[t]
        \centering
        \includegraphics[width = 0.95 \linewidth]{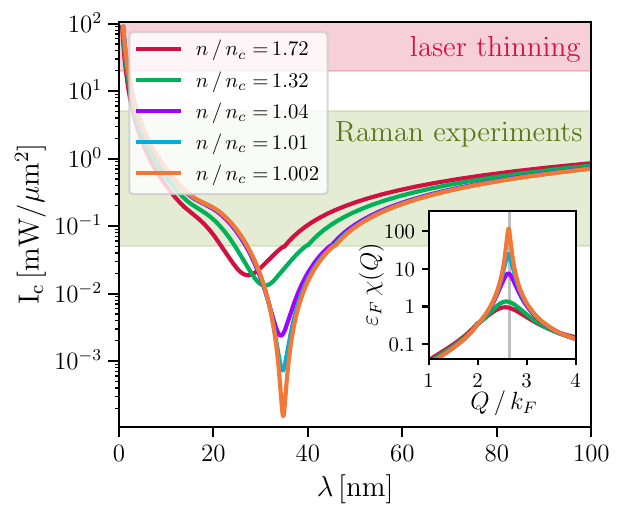}
        \caption{\textbf{Critical power near the Wigner crystal phase transition.} Near a critical density $n_c$, the enhanced static structure factor (inset) significantly reduces the critical pump intensity $\mathrm{I}_c$ if the grating period $\lambda$ matches the wavevector of the yet-unformed crystal. We use QMC data at zero temperature to compute $\chi(\mathbf{Q})$ and superimpose our results with two experimentally relevant regimes (shaded regions).  ``Laser thinning" refers to the regime wherein the onset of sample damage is expected~\cite{Hu2017MoS2LaserThinning}. The green region marks the intensity band of current Raman experiments in 2D materials~\cite{aslan_probing_2018,byrley_photochemically_2019}. The vertical line in the inset marks the ordering wavevector of the Wigner crystal. The dramatic decrease in threshold intensity highlights the importance of coupling to enhanced fluctuations.}
        \label{fig:qmc_plot} 
\end{figure}

Calculating the density response of an electronic fluid at low densities is not feasible within any straightforward perturbative scheme such as the random phase approximation (RPA).  This is because the electronic fluid is not perturbatively close to a simple metal but is rather a quantum melted electronic solid.  Instead, we use interpolated zero-temperature QMC data~\cite{PhysRevB.68.195210} to estimate the critical pump intensity as a function of the grating period at different electronic densities $n$ (Fig.~\ref{fig:qmc_plot}). 

Before articulating our quantitative results, we note that a comprehensive accounting of the experimental parameters used to model the excitonic, cavity, and substrate aspects of the problem is found in Section C of the Methods. We briefly summarize some of the qualitative conclusions. As polaron linewidths and the cavity leakage limit the smallest possible detunings $\Delta_A$ and $\Delta_{\rm c}$, the lowest sCDW thresholds will be achievable with high-quality samples possessing low disorder and long excitonic---thereby polaronic---lifetimes, embedded in 
optical cavities. Additionally, the grating material should have a high index of refraction to optimize the evanescent field strength (see SI), and the illuminated sample area should be maximized so as to leverage the collective  coupling from as many superradiant emitters as possible~\cite{coldAtoms1}. 

We now turn to our quantitative results. We find that $\chi(\mathbf{Q})$ diverges as the electronic density is lowered below $n_{c}$. The divergence occurs at the wavevector $Q=\sqrt{\frac{4\pi}{\sqrt{3}}}k_{\rm F}\approx 2.7k_{\rm F}$ commensurate with the yet-unformed Wigner crystal. As implied by Eq.~\eqref{eq:threshold}, it is optimal to choose $\lambda$ matching the period of the proximate Wigner crystal.
The predicted values of critical pump intensity $\mathrm{I}_c$ for the onset of superradiance suggest that continuous wave control is feasible. Estimates generally place $\mathrm{I}_c$ within the range of current Raman experiments~\cite{aslan_probing_2018,byrley_photochemically_2019}---a relevant comparison because they are high-power, continuous-wave equilibrium probes---dipping far below $2\times10^{-2}\rm{mW/\mu m^2}$ when $n$ is tuned withing $4\%$ of $n_c$. As shown in Fig.~\ref{fig:qmc_plot}, tuning the grating wavelength close to the periodicity of the proximate Wigner crystal can significantly lower $\mathrm{I}_c$. Note that the values quoted for the threshold intensity $\mathrm{I}_c$ account for the $\eta_{\mathbf{G}}\sim 10 \%$ conversion efficiency of the incident pump field into grating-wavevector component.
 
 We underscore, however, that our proposal does not yield a ``superradiant Wigner crystal." For example, we do not expect to photo-induce a triangular lattice of electronic density. That which we photo-induce is a \textit{superradiant charge density wave} with charge order at wave-vector $\mathbf{Q}$ induced by the interference between the electric field of the superradiant cavity and the near-field of the nanoscale grating (putatively, a stripe order). We further note that harnessing the enhanced fluctuations close to a density-ordered phase is a broadly applicable strategy that can be easily adapted for thermal phase transitions. We note in passing that there is controversy regarding the precise nature of the transition between the strongly correlated electronic liquid to Wigner crystalline solid takes place~\cite{wignercontroversy1,wignercontroversy2, wignercontroversy3, wignercontroversy4}. As such, quantitative details of the density response at particular values of the density may vary, but the qualitative fact that density fluctuations are strongly enhanced near the transition is clear.

\subsection{Finite temperature effects}\label{sec:rpa}

In the previous section, we highlighted how enhanced quantum fluctuations near a zero-temperature, quantum density-ordering transition can dramatically reduce the  laser power required to produce sCDWs. However, this necessitates accessing  temperature scales that are sub-$ 4~{\rm K}$, and this might prove challenging for early experimental implementations due to constraints on sample cooling power or pump laser heating. In monolayer TMDs, accessing the thermal melting transition---around which density fluctuations are also expected to diverge---occurs at higher temperatures ${\sim}10~{\rm K}$~\cite{smolenski_signatures_2021}, which is already within the temperature range of TMD-multimode cavity QED experiments~\cite{Hiller2026acv}. However, even far away from the Wigner crystal phase, density fluctuations are enhanced near nesting wave-vectors for the Fermi surface: Recall, e.g., the Kohn anomaly in 1D where density fluctuations diverge at $2k_{\rm F}$. While these fluctuations are more modest in higher dimensions, we show that they can be used to keep the superradiant threshold power far below the damage threshold of the TMD material. For readers familiar with the motivating cavity-atomic gas experiments this should not prove surprising: The experiments that initially demonstrated what was later identified as superradiance from a nonequilibrium Hepp-Lieb-Dicke transition were not performed on quantum degenerate gases but rather on thermal atomic clouds~\cite{coldAtoms1}. 

\begin{figure}[t!]
        \centering
        \includegraphics[width = 0.95 \linewidth]{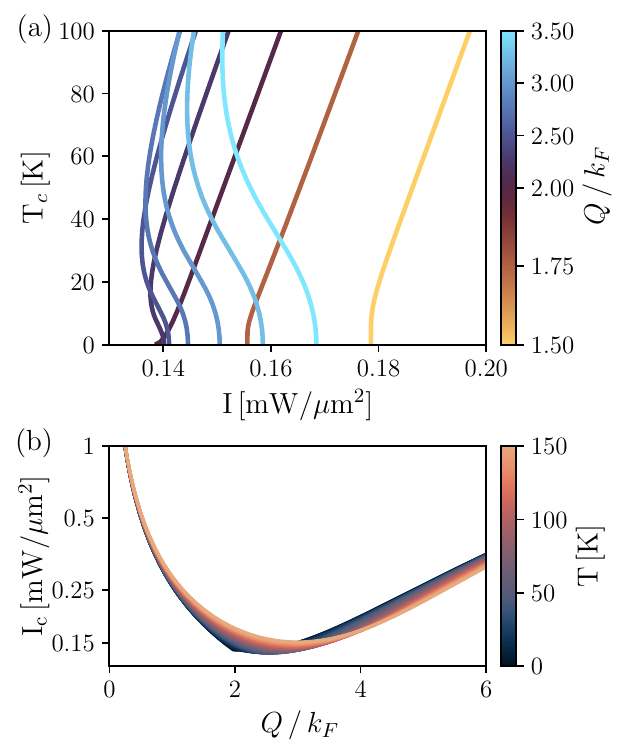}
        \caption{\textbf{Finite-temperature phase boundary.} (a) Critical temperature as a function of pump laser intensity for different ordering wavevectors $Q$, measured in units of the Fermi wavevector. Results were obtained through RPA calculations for $\varepsilon_F=7\rm~meV$. For $Q>2k_{\rm F}$, the phase boundaries imply re-entrant behavior upon cooling. (b) Critical pump intensity as a function of ordering momentum for a range of temperatures. To achieve the minimal laser power necessary for superradiance, it is optimal to tune the grating periodicity close to $2k_{\rm F}$.}
        \label{fig:rpa_linear_plot} 
        
\end{figure}

We substantiate the claim that sCDWs can be formed at practical experimental temperatures by performing a quantitative microscopic calculation at higher electronic densities where the density response of interacting electronic gases can be treated within RPA~\cite{DasSarma2004,MALDAGUE1978296}. Within these calculations, the peak in $\chi(\mathbf{Q})$ shifts from $Q=2.7k_{\rm F}$ to the Fermi surface diameter $2k_{\rm F}$. We find that even for moderately high-temperature electrons far from the Wigner crystal phase, the superradiant threshold may lie within the experimentally accessible regime (Fig.~\ref{fig:rpa_linear_plot}). In this high-density regime, it is optimal to tune the grating periodicity close to the wavelength matching the Fermi surface diameter $\pi/k_F$, which is approximately $10 ~\mathrm{nm}$ for the data shown~\cite{colormaps}. On closer inspection, Fig.~\ref{fig:rpa_linear_plot} reveals a striking dependence of the phase boundary on temperature: For gratings with $Q>k_F$, the electrons enter the superradiant phase and later exit upon cooling at a fixed pump intensity. As the grating period shrinks, the re-entrant region expands; see Fig.~\ref{fig:rpa_linear_plot}(a). This effect is due to thermal broadening of $\chi(\mathbf{Q},T)$ which has a non-monotonic dependence on temperature for large momenta in two-dimensional electron gases; see Fig.~\ref{fig:rpa_linear_plot}(b). We note that similar re-entrant superradiant order was discussed in the context of thermal bosons in Ref.~\cite{Piazza2013a}.

 \section{Discussion}
In this work, we have articulated a novel strategy for creating a photo-induced, superradiant charge density wave transition in solid-state systems. Our approach merges tools from nanophotonics (nanoscale gratings) with phenomena from excitonic physics (exciton-polarons) to construct a microscopic mechanism that enables electronic density order to strongly couple to an optical cavity driven using experimentally modest laser powers. We provide several design principles towards achieving lower laser powers for our proposal, in particular emphasizing that the nanoscale grating periodicity should match the length scale at which electronic density fluctuations peak. A particularly promising and generalizable strategy is to operate near quantum or thermal phase transitions. Our approach to continous wave optical control of electronic order by combining cavity enhancement with a low power laser drive is complementary to theoretical proposals aiming to sculpt superradiant phases of quantum matter in the solid state with only strongly coupled equilibrium vacuum or thermal fluctuations~\cite{cavssr1, cavssr2, cavssr3, cavssr4, cavssr5, cavssr6, cavssr7}.

In our work, we focused on continuous-wave optical control. Were the required laser powers to cause excessive heating for a particular setup, a pulsed laser may be used instead, provided that the duration of the pulse is minimally longer than the timescale of the dynamical instability of the coupled light-matter system following a quench across the superradiant phase transition~\cite{marijanović2024dynamicalinstabilitiesstronglyinteracting}\footnote{With the caveat that the photoinduced state should be experimentally characterizable, even if optically}. We estimate the instability growth rate for weak quenches to be approximately $\Delta_{\rm c}$ (see SI), finding that for our proposed setup pulse durations considerably longer than, but on the order of, $10 ~\rm ns$ would be sufficient to observe at least \textit{transient} sCDW order. 

The use of a multimode optical cavity in, e.g., the confocal geometry~\cite{Vaidya2018tpa}, may be employed to enhance coupling to cavity photons by engineering ``supermodes'' formed from a sum of degenerate cavity modes.  This substantial enhancement can compensate any reduction in cavity coupling due to the length of mirror spacing required to accommodate the TMD~\cite{confocal2,Bourzutschky2024rci} and the nanofabricated substrate. Our single-mode considerations here can be straightforwardly extended to multiple cavity modes~\cite{PhysRevX.15.021089, marijanović2024dynamicalinstabilitiesstronglyinteracting}.

Beyond monolayer materials, our coupling scheme can be extended to moir\'{e} systems with moir\'{e} excitons~\cite{Xiong2023}. Twisted layered materials have intrinsically large electronic length scales that could enable efficient coupling to the light scattered from the grating. By applying a magnetic field to imbalance valley populations, one could intertwine valley physics with superradiance.

We close this work by speculating on the experimental phenomenology of sCDWs. Most notably, we expect that sCDWs will exhibit strikingly distinct transport phenomenology from plain-vanilla charge density waves. In conventional CDWs, disorder-induced pinning of the phase degree of freedom is known to govern the transport phenomenology. Pinning leads to, for example, conductivity thresholds in the DC transport and a pinning mode emerging in the AC transport~\cite{Gruner1994}. In stark contrast to conventional charge density waves, the quantum-optical origin of sCDW order unusually couples the local phase of the order parameter to the amplitude of the charge order. Sufficiently deep in the superradiant phase, we suspect that pinning domains may not exist, rendering the phenomenology of sCDWs drastically different from CDWs but instead similar to that of systems in which macroscopic phase coherence plays a key role in transport, such as Josephson junctions. A crucial modification will be that the aforementioned macroscopic phase coherence of the charge order will be coupled to the driven cavity mode. A careful examination of the coupled transport and optical phenomenology expected in such a system is the subject of a paper in preparation by the authors. 
\newline

\section{Methods}

\subsection{Deriving the Effective Hamiltonian}

We derive the effective electron-photon coupling used in the main text, starting from the microscopic Hamiltonian introduced in the Results section,
\begin{widetext}
\begin{align}
H =& \sum\limits_{\mathbf{p}, \sigma} \varepsilon^\sigma_\mathbf{p} c^\dag_{\mathbf{p}, \sigma} c_{\mathbf{p}, \sigma} + \!\!\sum \limits_{\substack{\mathbf{p}, \mathbf{p}', \mathbf{q},\sigma,\sigma'}} \! V_{\mathbf{q}} c^\dag_{\mathbf{p} + \mathbf{q},\sigma} c^\dag_{\mathbf{p}'- \mathbf{q}, \sigma'} c_{\mathbf{p}', \sigma'} c_{\mathbf{p},\sigma} + \!\! \sum \limits_{\substack{\mathbf{p}, \mathbf{p}', \mathbf{q}, \sigma, \sigma'}} \!   U^{\sigma,\sigma'}_{\mathbf{q}} X^\dag_{\mathbf{p} + \mathbf{q},\sigma} c^\dag_{\mathbf{p}' -\mathbf{q}, \sigma'} c_{\mathbf{p}', \sigma'} X_{\mathbf{p}, \sigma}+\nonumber\\
&+ \sum_{\mathbf{k},\sigma} \Omega_\mathbf{k} X^\dagger_{\mathbf{k},\sigma} X_{\mathbf{k},\sigma}
+ d_x \sum_{\mathbf{k},\sigma} E_\mathbf{k} (X_{-\mathbf{k},\sigma} + X_{\mathbf{k},\sigma}^\dag)
+ \omega_c a^\dagger a,\label{eq:H_eff_expand}
\end{align}
\end{widetext}
which describes electrons coupled to excitons and a cavity photon mode. Here $c^\dagger_{\mathbf{k},\sigma}$ creates an electron with momentum $\mathbf{k}$ and valley index $\sigma$, and $X^\dagger_{\mathbf{k},\sigma}$ creates an exciton with center-of-mass momentum $\mathbf{k}$ and dispersion $\Omega_\mathbf{k}$.

As discussed in the Results section, the electric field entering Eq.~\eqref{eq:H_full} contains both the classical pump field and the quantized cavity field. The nanoscale grating converts the near-infrared pump beam into evanescent components with large in-plane momentum $\mathbf{G}$, so that
\begin{widetext}
\begin{align}
E_\mathbf{k} =
E_G (e^{-i\omega_p t} +e^{i\omega_p t})(\delta_{\mathbf{k}, \mathbf{G}}
+\delta_{\mathbf{k},-\mathbf{G}})+E_{\rm c} (a + a^\dagger) ( \delta_{\mathbf{k}, + \mathbf{k}_{\rm c}}+ \delta_{\mathbf{k}, -\mathbf{k}_{\rm c}}),
\end{align}
\end{widetext}
where $E_G$ is the amplitude of the grating-generated pump field and $E_{\rm c}$ is the vacuum electric field amplitude of the cavity mode.

In the presence of a Fermi sea, excitons hybridize with particle-hole excitations of the electron gas and form exciton-polaron quasiparticles~\cite{sidler2017fermi}. We focus on the attractive-polaron branch, which corresponds to the lowest-energy excitonic excitation at finite electron density. The attractive-polaron state with total momentum $\mathbf{Q}$ may be written using the Chevy variational ansatz~\cite{sidler2017fermi}

\begin{widetext}
\begin{align}
\ket{A_{\mathbf{Q},\sigma}} =
\sqrt{Z_\mathbf{Q}}\,X^\dagger_{\mathbf{Q},\sigma}|0\rangle
+
\sqrt{1-Z_{\mathbf{Q}}}
\sum_{q>k_F,\,p<k_F}
\phi_{\mathbf{p}-\mathbf{q}}\,
X^\dagger_{\mathbf{Q}+ \mathbf{p}- \mathbf{q},\sigma}
c^\dagger_{\mathbf{q},\overline{\sigma}}c_{\mathbf{p},\overline{\sigma}}\ket{0},\label{eq:chevy}
\end{align}
\end{widetext}
where $|0\rangle$ is the tensor product of the electronic ground state with the exciton vacuum. The first term in the wavefunction \eqref{eq:chevy} corresponds to a bare exciton, while the second term describes an exciton dressed by particle-hole excitations of the Fermi sea. The parameter $Z_\mathbf{Q}$ is the quasiparticle residue of the attractive polaron and determines the weight of the bare-exciton component. The relative wavefunction $\phi_\mathbf{q}$ describes the three-body electron-exciton state and can be found variationally for given potentials $V$ and $U$ \cite{sidler2017fermi,wang2025spectroscopywignercrystalpolarons}. It is normalized as
\begin{equation}
Z_\mathbf{Q} + (1-Z_\mathbf{Q})\sum_\mathbf{q} |\phi_\mathbf{q}|^2 = 1 .
\end{equation}

The light-matter interaction term $H_{LM} =
d_x \sum\limits_{\mathbf{k},\sigma}
E_\mathbf{k} (X_{-\mathbf{k},\sigma} + X_{\mathbf{k},\sigma}^\dag)$ creates or annihilates single excitons, which means it will couple subspaces with zero and single polarons. We take $\ket{A_{\mathbf{Q},\sigma}}$ to be the approximate eigenstates of the system. Since we will consider only virtual excitations of polarons, it is sufficient to project the full Hamiltonian \eqref{eq:H_eff_expand} onto only the zero- and single-polaron sectors. To that end, we introduce corresponding projectors 
\begin{equation}
P = \mathbbm{1}_e \otimes |0_x\rangle\langle0_x|,
\qquad
Q = \sum_{\mathbf{Q},\sigma} |A_{\mathbf{Q},\sigma}\rangle\langle A_{\mathbf{Q},\sigma}|.
\end{equation}
Projected onto the zero- and single-polaron subspaces, the effective Hamiltonian reads
\begin{align}
    H_{\rm eff} = H_e+\sum_{\mathbf{Q},\sigma} E^{(A)}_\mathbf{Q} |A_{\mathbf{Q},\sigma}\rangle\langle A_{\mathbf{Q},\sigma}| + H_{LM}^{\mathrm{proj}} + \omega_c a^\dagger a,\label{eq:Heff_proj}
\end{align}
where $E^{(A)}_\mathbf{Q}$ is the energy dispersion of the attractive polaron. The term $H_{LM}^{\mathrm{proj}}$ captures the light-driven transitions between the zero- and single-polaron sectors, with the strength of the transition given by the matrix element $\bra{0_x}X_{\mathbf{k},\sigma}\ket{A_{\mathbf{Q},\sigma}}$ of the electric dipole term. To obtain a low-energy description of the coupled electron-photon system, we first move to a frame rotating at the pump frequency $\omega_p$ and perform the rotating-wave approximation. The largest scale left in the problem is the polaron-pump detuning $\Delta_\mathbf{Q} = E_\mathbf{Q}^{(A)}-\omega_p$, which allows us to integrate out the \textit{virtually excited} attractive-polaron states in the dispersive limit. The effective Hamiltonian acting within the low-energy electronic sector is
\begin{equation}
H_{\mathrm{eff}}=PHP + PHQ\frac{1}{E-QHQ} QHP .\label{eq:Heff_general}
\end{equation}
The first term $PHP=H_e + \Delta_c a^\dagger a$ contains the electronic Hamiltonian and the transformed cavity photon term with the cavity-pump detuning $\Delta_c = \omega_c-\omega_p$. Since the polaron-pump detuning $\Delta_\mathbf{Q}$ is larger than any energy scale in the effective Hamiltonian, the resolvent may be approximated as
\begin{equation}
(E-QHQ)^{-1} \simeq -\frac{1}{\Delta_\mathbf{Q}}.\label{eq:resolvent}
\end{equation}
In evaluating the second term in Eq.~\eqref{eq:Heff_general}, we consider processes in which a pump photon creates a virtual attractive polaron that subsequently dissolves by emitting a cavity photon, see Fig. \ref{fig:setup}(c). We also consider processes in which the virtual polaron is created by absorbing a cavity photon and annihilated by emitting into the pump field. The difference of photon momenta $\pm \mathbf{G}\pm\mathbf{k}_{\rm c}$ is transferred to the electronic system. The resulting interaction couples the cavity field to electronic density fluctuations,

\begin{widetext}
\begin{align}
    H^{\mathrm{eff}}_{LM} = -d_x^2 E_\mathbf{G} E_{\rm c} \sum\limits_{\mathbf{Q}, \mathbf{q},\sigma} \left(a^\dag \frac{\sqrt{ Z_{\mathbf{G}} (1-Z_{\mathbf{G}})}}{\Delta_\mathbf{G}} + a \frac{ \sqrt{ Z_{\mathbf{k}_{\rm c}} (1-Z_{ \mathbf{k}_{\rm c}} )}}{\Delta_{ \mathbf{k}_{\rm c}}} \right) \phi_{ \mathbf{Q}} c^\dag_{\mathbf{q}+ \mathbf{Q}, \sigma} c_{\mathbf{q}, \sigma}  +h.c.\label{eq:method_Heff}
\end{align}
\end{widetext}
where $\mathbf{Q} = \pm \mathbf{G} \pm \mathbf{k}_c$. 

Introducing the electronic density operator $\rho_\mathbf{Q} =
\sum_{\mathbf{q},\sigma} c^\dagger_{\mathbf{q}+\mathbf{Q},\sigma}c_{\mathbf{q},\sigma}$, the effective Hamiltonian \eqref{eq:method_Heff} may be written in the compact form \eqref{eq:H_eff}. We now elaborate on all the factors entering the light-matter coupling strength in \eqref{eq:method_Heff}. The coupling originates from interference between the two components of the attractive-polaron wavefunction \eqref{eq:chevy}: when the virtual polaron has total momentum $\mathbf{G}$, the pump field couples to the bare-exciton component with amplitude $\sqrt{Z_\mathbf{G}}$. On the other hand, the return process that emits a cavity photon and generates the electronic density operator couples to the particle-hole-dressed component with amplitude $\sqrt{1-Z_\mathbf{G}}$. Altogether, the combined processes produce the factor $\sqrt{Z_\mathbf{G}(1-Z_\mathbf{G})}$ controlling the strength of the effective electron-photon interaction and the factor $\Delta_{\mathbf{G}}^{-1}$ comes from the resolvent \eqref{eq:resolvent}. The wavefunction $\phi_\mathbf{Q}$ carries information about the momentum transfer between electrons. A similar argument holds when the cavity photon is absorbed to create a virtually excited polaron with total momentum $\mathbf{k}_{\rm c}$, generating the $\sqrt{Z_{\mathbf{k}_{\rm c}}(1-Z_{\mathbf{k}_{\rm c}})}$ term. The two processes we just described and their time-reversed counterparts, visually represented by connecting the pump and cavity lines in Fig. \ref{fig:setup}(c), produce all the terms in Eq.~\eqref{eq:method_Heff}.

Since the wavevectors of the pump and cavity fields are much smaller than trionic and exitonic length scales, we can approximate $Z_\mathbf{G}$ and $Z_{\mathbf{k}_{\rm c}}$ with the quasiparticle residue $Z$ for polarons that have zero total momentum. Working with electronic systems where $k_Fa_T\ll 1$, we use the known result $Z=\frac{m_e}{\mu_T}\frac{\varepsilon_F}{\varepsilon_T} = k_{\rm F}^2a_T^2$ and consequently $1-Z\approx 1$~\cite{PhysRevB.95.035417}. Furthermore, since holes can only carry momenta smaller than $k_F$, the wavefunction $\phi_\mathbf{q}$ can be simplified to the bound trion wavefunction in the dilute limit of the attractive polaron where the small $k_F$ sets the ultraviolet cut-off for the electronic (hole) degrees of freedom. In that limit, $\mathbf{q} $ is the relative momentum of the electron and exciton. By plugging in $\phi_{\mathbf{q}}=\frac{4\pi a_T}{k_F A}\Theta(a_T^{-1}-q)$ that is cut off at the trion binding radius, and assuming the polaron dispersion to be much smaller than the polaron-pump detuning, i.e. $\vert\Delta_{\mathbf{G}} - \Delta_{\mathbf{k}_{\rm c}}\vert \ll \Delta_A$, we find the light-matter coupling strength in Eq.~\eqref{eq:lambda}.

We note that we consider a specific set of light-matter interactions that contribute to ordering at $\mathbf{Q}_\pm = \mathbf{G}\pm \mathbf{k}_{\rm c}$. We briefly discuss why we do not consider other terms. Contributions arising from the interference between the grating-scattered cavity field components at $\pm \mathbf{G}\pm\mathbf{k}_{\rm c}$ and the homogenous pump term interfere constructively with contributions that we consider. Contributions arising from the interference between the  cavity field components at $\mathbf{k}_{\rm c}$ with the homogenous pump term induce density response at $\mathbf{Q} \approx \mathbf{k}_{\rm c}$. The response here in a charged electronic fluid is negligible as $\chi(\mathbf{k}_{\rm c}) \ll \chi(\mathbf{Q})$. Finally, ``pump-pump" interference between the pump evanescent component at momentum $\mathbf{G}$ and the homogenous field lead to an optical lattice effect which restructures the homogenous electronic gas. This restruturing is not connected to the superradiant transition but it does enhance the fluctuations near momenta $\mathbf{Q_\pm}$. Accordingly, while we do not explicitly consider the role of these effects within our work for clarity, we believe that each of these terms will actually lower the threshold for superradiant ordering.

\subsection{Near-field simulations of the grating}\label{app_grating}
We perform two-dimensional frequency-domain simulations using the finite-element method via the COMSOL software package to calculate the electric field distribution directly above the nano-grating structure, thereby estimating the exciton density wave-function generated in semiconductor monolayers under optical excitation. Periodic boundary conditions are applied to model the device’s periodic perturbations. The structure is illuminated by a normally incident plane wave with the electric field polarized perpendicular to the nano-grating, with all modulated spatial modes of the evanescent field included. Three different substrate materials are considered to investigate their impact on near-field modulation, with the refractive indices of \ce{Ta2O5}, \ce{Al2O3}, and \ce{SiO2} set to 2.10, 1.76, and 1.45, respectively (see SI). The refractive index of hexagonal boron nitride, which functions as the spacer between the semiconductor monolayer and the nano-grating, is taken to be 2.10. The optimal efficiency of evanescent field coupling $\eta_{\mathbf{G}}$, defined as the ratio of the intensity of the first-order spatial mode of the short-wavelength, evanescent electric field versus the total intensity at the TMD's position, was found to be 0.12 for \ce{Ta2O5}.

\subsection{Experimental Parameters}
In preparing Figs.~\ref{fig:qmc_plot} and \ref{fig:rpa_linear_plot}, we took parameter values $\epsilon=4.5$ for encapsulation in hBN and $m_e=0.5m_0$ for the effective electron mass in $\mathrm{MoSe}_2$. The exciton dipole moment $d_{\rm x}=\mu_{cv}\psi_x(0)\sqrt{A}$, where $\mu_{cv}$ is the interband transition matrix element and $\psi_x$ is the exciton wavefunction in position space, was estimated by $\mu_{cv}\psi_x(0) \approx 0.9 e_0$ from excitons' radiative lifetimes~\cite{Atac, ExcitonLinewidth}. The linewidth of the attractive polaron $\Gamma_A=5\mathrm{~meV}$, and correspondingly $\Delta_A=2\Gamma_A$~\cite{sidler2017fermi, PhysRevB.108.125406}. The trion binding energy $\varepsilon_T=25\mathrm{~meV}$ and trion radius $a_{T}=2.1\rm ~nm$ were taken from~\cite{PhysRevB.95.081301,sidler2017fermi}. The cavity mode volume $V_{\rm c}=\pi (w_0/2)^2 L$ was estimated for a cavity mode waist $w_0=10\mathrm{~\mu m}$ and cavity length $L=1\mathrm{~cm}$; waists down to $<$2~$\mu$m are possible using multimode cavities~\cite{confocal2}. In addition, $ A=100\mathrm{~\mu m}^2$ was assumed for the sample area, $\eta_{\mathbf{G}}=0.12$ for the efficiency of evanescent field scattering, $\lambda_{\rm c}=750\rm{~nm}$ for the cavity mode wavelength and $\Delta_{\rm c}=100\mathrm{~MHz}$ for the cavity-pump detuning in a high-finesse optical cavity.

\subsection{Dissipative cavities}\label{app_dissipation}
In practice, any cavity has a finite field emission rate $\kappa$. Applying input-output theory, we incorporate photon loss by changing $\frac{\mathrm{d} \hat{A}}{\mathrm{d}t} \Rightarrow \frac{\mathrm{d} \hat{A}}{\mathrm{d}t}-\frac{\kappa}{2}\hat{A}$ for $\hat{A}=\hat{X}_{\rm c},\, \hat{P}_{\rm c}$ which implies $\omega \hat{A}= \comm{\hat{H}}{\hat{A}} + i\frac{\kappa}{2}\hat{A}$~\cite{collett_squeezing_1984}. The eigenmodes and instabilities of the system shift in frequency
\begin{equation}
    \Delta_{\rm c}^2 = \left(\omega-i\frac{\kappa}{2}\right)^2 + 2\Delta_{\rm c} N_{\rm e} \Lambda^2 \sum\limits_\mathbf{Q} \chi(\mathbf{Q}, \omega),
\end{equation}
and the instability threshold is raised by
\begin{equation}
    \Delta_{\rm c}^2\left(1+\frac{1}{4}\left( \frac{\kappa}{\Delta_{\rm c}}\right)^2\right) = 2\Delta_{\rm c} N_{\rm e} \Lambda^2 \sum\limits_\mathbf{Q} \chi(\mathbf{Q}).
\end{equation}
The cavity loss rate does not affect $\chi(\mathbf{Q},\omega)$, since $\chi$ is a property of the electronic system alone. We cast the equation in a convenient form
\begin{equation}
    \Lambda_c^2 = \frac{\varepsilon_F \Tilde{\Delta}_{\rm c}^2}{4 N_e\alpha\Delta_{\rm c}}= \frac{ \pi \hbar^2 \Tilde{\Delta}_{\rm c}^2}{4 A\alpha\Delta_{\rm c} m^*},
\end{equation}
where $\Tilde{\Delta}_{\rm c}=\sqrt{\Delta_{\rm c}^2+(\kappa/2)^2}$, $\alpha=\varepsilon_F \chi(\mathbf{Q})$ and we assumed $\chi(\mathbf{Q}_\pm)\approx\chi(\mathbf{Q})$.


\subsection{RPA response functions}
\label{app_finiteT}
Recall the Lindhard function for static density susceptibility for zero-temperature, non-interacting electrons~\cite{mihaila2011lindhardfunctionddimensionalfermi}
\begin{equation}
    \chi_0(\mathbf{Q}, T=0)= \frac{1}{\varepsilon_F}\left(1- \Theta(Q -2k_{\rm F})\sqrt{1-(2k_{\rm F}/Q)^2}\right),
\end{equation}
where $\Theta(Q)$ is the step function. At a finite temperature $T$ and a fixed number of electrons, the non-interacting response function is~\cite{DasSarma2004, MALDAGUE1978296}
\begin{equation}
    \chi_0(\mathbf{Q},T) = \int\limits_0^\infty \mathrm{d}\mu' \frac{\chi_0(\mathbf{Q},\mu', T=0)}{4T\mathrm{cosh}^2 \left(\frac{\mu(T) -\mu'}{2T}\right)},
\end{equation}
where $\chi_0(\mathbf{Q},\mu', T=0)$ is the zero-temperature non-interacting susceptibility with Fermi energy $\mu'$ and
\begin{equation}
    \mu(T) = T\,\mathrm{ln}\left( e^{\varepsilon_F/ T} - 1\right)
\end{equation}
is the temperature-dependent chemical potential that fixes the number of electrons. To include repulsion between electrons through RPA, we define
\begin{equation}
    \chi^{\rm RPA}(\mathbf{Q},T) = \frac{\chi_0(\mathbf{Q},T)}{1 + \nu_{\rm e}v_\mathbf{Q}\chi_0(\mathbf{Q},T)},
\end{equation}
where $\nu_{\rm e}=\frac{k_{\rm F}^2}{2\pi}$ is the two-dimensional electron density and $v_\mathbf{Q}=\int \mathrm{d}^2 \mathbf{r} v(\mathbf{r})e^{-i\mathbf{Q}\cdot\mathbf{r}}= \frac{e_0^2 2\pi}{\varepsilon Q}$ is the Fourier transform of the Coulomb potential, which we take as an approximation of electron-electron interactions.

\bibliography{bibliography}

\section{Acknowledgments}
We acknowledge stimulating discussions with Hadyn Samuel Adlong, Alex Gomez Salvador, Atac İmamoğlu, Tony Heinz, Mark Brongersma, Jennifer Dionne, Han Hiller, Pranav Parakh, Samuel Aronson and especially Ilya Esterlis and Arthur Christianen.

\section{Author contributions}
E.D., J.K., \& B.L.L. concieved and supervised the project. L.S., S.C., F.M. and E.D. developed the theoretical framework. L.S., S.C. and Q.L. performed the numerical computations. All authors participated in the discussion and interpretation of results. L.S., S.C. and F.M. wrote the manuscript with input from all authors.

\section{Funding}
J.K. and B.L.L. acknowledge funding from the Gordon and Betty Moore Foundation (grant number GBMF10693). The ETH group acknowledges funding from the SNSF project 200021\textunderscore212899, 
the SNSF Sinergia grant CRSII--222792 and
the Swiss State Secretariat for Education, Research and Innovation (contract number UeM019-1).



\onecolumngrid
\appendix
\newpage
\begin{center}
	\textbf{\Large Supplementary Materials}
\end{center}
\normalsize

\setcounter{equation}{0}
\setcounter{figure}{0}
\setcounter{table}{0}
\renewcommand{\thefigure}{S\arabic{figure}}
\makeatletter
\setlength\tabcolsep{10pt}
\setcounter{secnumdepth}{2}


\section{Choice of substrate material}\label{app_substrate_material}
\begin{figure}[h!]
        \centering
        \includegraphics[width = 0.95 \linewidth]{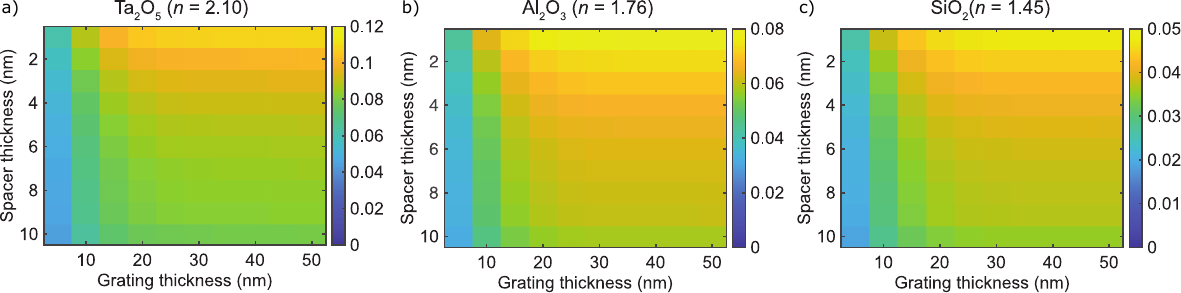}
        \caption{\textbf{Scattering efficiency} $\eta_{\mathbf{G}}$ defined as the ratio of the intensity due to the first-order spatial mode component to the intensity of the long-wavelength laser field.}
        \label{fig:tmd_efficiency} 
        
\end{figure}

As explained in the Methods section, we performed finite-element simulations (using COMSOL) for three different materials. We computed $\eta_{\mathbf{G}}$, the efficiency of scattering into the evanescent fields, defined as the ratio of the intensity due to the first-order spatial mode of the modulated, short-wavelength, evanescent electric field versus the total intensity at the TMD's position. Figure~\ref{fig:tmd_efficiency} shows that the higher the index of refraction of the substrate material and the closer the TMD is placed to the substrate (thinner spacer), the higher the efficiency. This can be understood intuitively from the evanescent nature of the high-momentum field component that decays exponentially as the spacer thickness increases.

\section{Self-consistency equation}\label{app_RPA}
We derive the self-consistency equation \eqref{eq:self_consistency} from the effective model \eqref{eq:H_eff} through RPA. For clarity, we distinguish operators $\hat{A}$ and expectation values $A$ in this section. We map the derivation onto a more general problem of linear response of electrons to external drives and use existing results from literature. Firstly, we notice that to close the system of equations found for $\hat{X}_{\rm c}$, $\hat{P}_{\rm c}$ and $\hat{\rho}_{\mathbf{Q}}$ in the main text, we need to compute the density operator $\hat{\rho}_{\mathbf{Q}}$ up to only linear order in $\hat{X}_{\rm c}$, since the expectation values of $\hat{X}_{\rm c}$ and $\hat{P}_{\rm c}$ vanish for $\Lambda=0$. In the framework of linearized equations of motion, we can thus take expectation values up to linear order in deviation from equilibrium at $\Lambda=0$. Consequently, there is no difference between taking expectation value of $\hat{X}_{\rm c}$ before or after computing the commutators $\comm{\hat{H}^{int}}{\hat{A}}$ within the RPA framework. This is a key insight, since it allows us to rewrite Eq.~\eqref{eq:H_eff} as
\begin{equation}
    \hat{H}^{int} = \Lambda \sum\limits_{\mathbf{Q}=\pm\mathbf{Q}_\pm}\left(X_{\rm c} \hat{\rho}_{\mathbf{Q}} + \hat{X}_{\rm c} \rho_{\mathbf{Q}}\right).\label{eq:H_int_app}
\end{equation}
In the way electronic density operators enter \eqref{eq:H_int_app}, we recognize the canonical linear response-type density driving term $\phi_{\mathbf{Q}} \hat{\rho}_\mathbf{Q}$. The commonly used field $\phi_\mathbf{Q}$ has been replaced by the expectation value of the photon displacement $X_{\rm c}$. We can apply results from linear response theory to state
\begin{equation}
    \rho_\mathbf{Q}(\mathbf{Q},\omega) = \chi(\mathbf{Q},\omega) N_{\rm e}\Lambda X_{\rm c}(\omega),
\end{equation}
where the equality holds for every $\mathbf{Q}$ component individually, up to linear order~\cite{dupuis2023field}. Therefore, the self-consistency equation \eqref{eq:self_consistency} holds generally. For interacting electrons, we need to use the $\chi(\mathbf{Q},\omega)$ computed in the interacting ground state.


\section{Instability timescale}
By quenching the coupling strength $\Lambda$ from 0 to a value above $\Lambda_c$, we induce a dynamical instability in which the electronic charge-density order and the photon displacement grow exponentially in time~\cite{marijanović2024dynamicalinstabilitiesstronglyinteracting}. We compute the growth rate $\Gamma=i\omega$ for weak quenches $\Lambda-\Lambda_c\ll\Lambda_c$, where $\Gamma\ll \varepsilon_F$, and we can simplify Eq.~\eqref{eq:self_consistency} to
\begin{equation}
    1+\left(\frac{\Gamma}{\Delta_{\rm c}}\right)^2 = 4\frac{\Lambda^2 N_{\rm e}\chi(\mathbf{Q})}{\Delta_{\rm c}},
\end{equation}
from which we can compute the growth rate 
\begin{equation}
    \Gamma = \Delta_{\rm c}\sqrt{2\frac{\Lambda - \Lambda_c}{\Lambda_c}}.
\end{equation}

\end{document}